\documentclass[preprint,showpacs,showkeys,aps,prb]{revtex4}
\usepackage[dvips]{graphicx}

\begin{document}

\title{Time-Reversible Random Number Generators : \\
       Solution of Our Challenge by Federico Ricci-Tersenghi \\
}

\author{
Wm. G. Hoover and Carol G. Hoover               \\
Ruby Valley Research Institute                  \\
Highway Contract 60, Box 601                    \\
Ruby Valley, Nevada 89833                       \\
}

\date{\today}



\vspace{0.1cm}

\begin{abstract}
Nearly all the evolution equations of physics are time-reversible, in the sense
that a movie of the solution, played backwards, would obey exactly the same
differential equations as the original forward solution.  By way of contrast,
{\it stochastic} approaches are typically {\it not} time-reversible, though they
could be  made so by the simple expedient of storing their underlying pseudorandom
numbers in an array.  Here we illustrate the notion of {\it time-reversible random
number generators}.  In Version 1 we offered a suitable reward for the first arXiv response
furnishing a reversed version of an only slightly-more-complicated pseudorandom
number generator.  Here we include Professor Ricci-Tersenghi's prize-winning
reversed version as described in his arXiv:1305.1805 contribution:
``The Solution to the Challenge in `Time-Reversible Random Number Generators' by
Wm. G. Hoover and Carol G. Hoover''.
\end{abstract}

\maketitle

\section{Time Reversibility}

The Newtonian, Lagrangian, or Hamiltonian microscopic motion equations ,
$$
\{ \ F = m\ddot r \ \} \ ; \ \{ \ p = (\partial {\cal L}/\partial \dot q) \ ; \
\dot p = (\partial {\cal L}/\partial \dot q) \ \} \ ; \ 
\{ \ \dot q = +(\partial {\cal H}/\partial \dot p) \ ; \ \dot p = -(\partial {\cal H}/\partial \dot q) \ \} \ ,
$$
even when embellished with thermostats, ergostats, or barostats, are typically time-reversible\cite{b1}.
The St$\phi$rmer-Verlet time-symmetric ``Leapfrog Algorithm'' :
$$
\{ \ q(t+dt) - 2q(t) + q(t-dt) \equiv (F(t)/m)dt^2 \ \} \ , 
$$
which can be iterated either forward or backward once the coordinates are given at two successive times,
is the most transparent example of time reversibility.  Likewise the Schr\"odinger equation and Maxwell's
electromagnetic field equations can be used to generate movies which obey exactly the same equations
whether projected in the ``forward'' or the ``backward'' direction of time.  Mathematicians
have considered more general definitions of time reversibility\cite{b2}, but in some cases these
generalizations would include the damped oscillator among the class of reversible systems\cite{b3},
which we believe isn't sensible.

The Langevin equation ,
$$
\{ \ m\ddot r = F - (m\dot r/\tau) + {\cal R} \ \} \ ,
$$
where both the drag coefficient $(1/\tau)$ and the random force ${\cal R}$ aren't time-reversible, is often used
in molecular simulations\cite{b4,b5}.  In order to make it possible to extend stochastic solutions both forward and
backward in time, and to simplify the reproducibility of numerical results by others we think it is desirable
to incorporate time-reversible random number generators in our otherwise deterministic algorithms.  The
following Section illustrates this idea with a simple example algorithm, too simple for serious use in
simulation.  The final Section of Version 1 challenged the reader to find a time-reversed version of a useful
algorithm, with a reward for being ``first'' with that specific algorithm.  In this Version we include the
successful algorithm found by Federico Ricci-Tersenghi {\it within a day} of the challenge's publication.

\section{An Oversimplified Pseudorandom Number Generator and its Time-Reversed Version}
The design of reversible random number algorithms can be illuminated by the study of time-reversible maps.
Kum and Hoover\cite{b6} considered two-dimensional maps which are time-reversible in the physicist's sense :
$$
M(+q,+p) = (+q',+p') \ ; \ M(+q',-p') = (+q,-p) \ .
$$
They pointed out that simple shears, with $\delta q \propto \delta p$ or $\delta p \propto \delta q$ , as well
as certain phase-space reflection operations, are time-reversible and can be combined with periodic boundary
conditions so that the points $(q,p)$ remain within the unit square.  In addition, if $Q$, $P$, and $R$ are
time-reversible maps then symmetric combinations of them, like $QPRPQ$ , are likewise time-reversible.  The
simple $q$ and $p$ shears very closely resemble typical algorithms for pseudorandom numbers, such as the
FORTRAN example function : \\

\vspace{5 mm}

{\tt
\noindent    
      function rund(intx,inty) \\
      i = 1029*intx + 1731 \\
      j = i + 1029*inty + 507*intx - 1731 \\
      intx = mod(i,2048) \\
      j = j + (i - intx)/2048 \\
      inty = mod(j,2048) \\
      rund = (intx + 2048*inty)/4194304.0 \\
      return \\
      end \\
}
\vspace{5 mm}

\noindent
This generator returns a periodic sequence of $2^{22}$ pseudorandom numbers, updating the two seed variables
{\tt intx} and {\tt inty} as it goes.  The least significant 11 binary digits of {\tt rund}'s numerator
can be generated by the simpler function : \\

\vspace{5 mm}

\noindent
{\tt
      function next(it) \\
      next = mod(1029*it + 1731,2048) \\
      return \\
      end \\
}

\vspace{5 mm}

\newpage

\noindent
The Figure at the bottom of this page is a plot of the 2048 iterates of {\tt next} , ordered according to
the index {\tt it} .  The first six entries are the pairs
$$
{\tt next} \rightarrow (1,712),(2,1741),(3,722),(4,1751),(5,732),(6,1761) \ .
$$
The single-valued plot of these forward iterates gives the (misleading) impression of {\it continuous}
lines with a slope (from either the odd or the even entries) of (5/1) , while the actual function is
{\it discontinuous} between successive entries due to the jumpy nature of the {\tt mod} function.
Reflecting the plot ,
$$
( \ x,y \ ) \longleftrightarrow ( \ y,x \ ) \ ,
$$
illustrates the output of the time-reversed algorithm {\tt last} .  Because the equations are {\it linear}
it is relatively easy to find the analytic form of the reversed function : \\

\vspace{5 mm}

\noindent
{\tt
      function last(j) \\
      last = mod(205*j + 1497,2048) \\
      return \\
      end \\
}

\begin{figure}[h]
\vspace{1 cm}
\includegraphics[width=8.90cm,angle=-90]{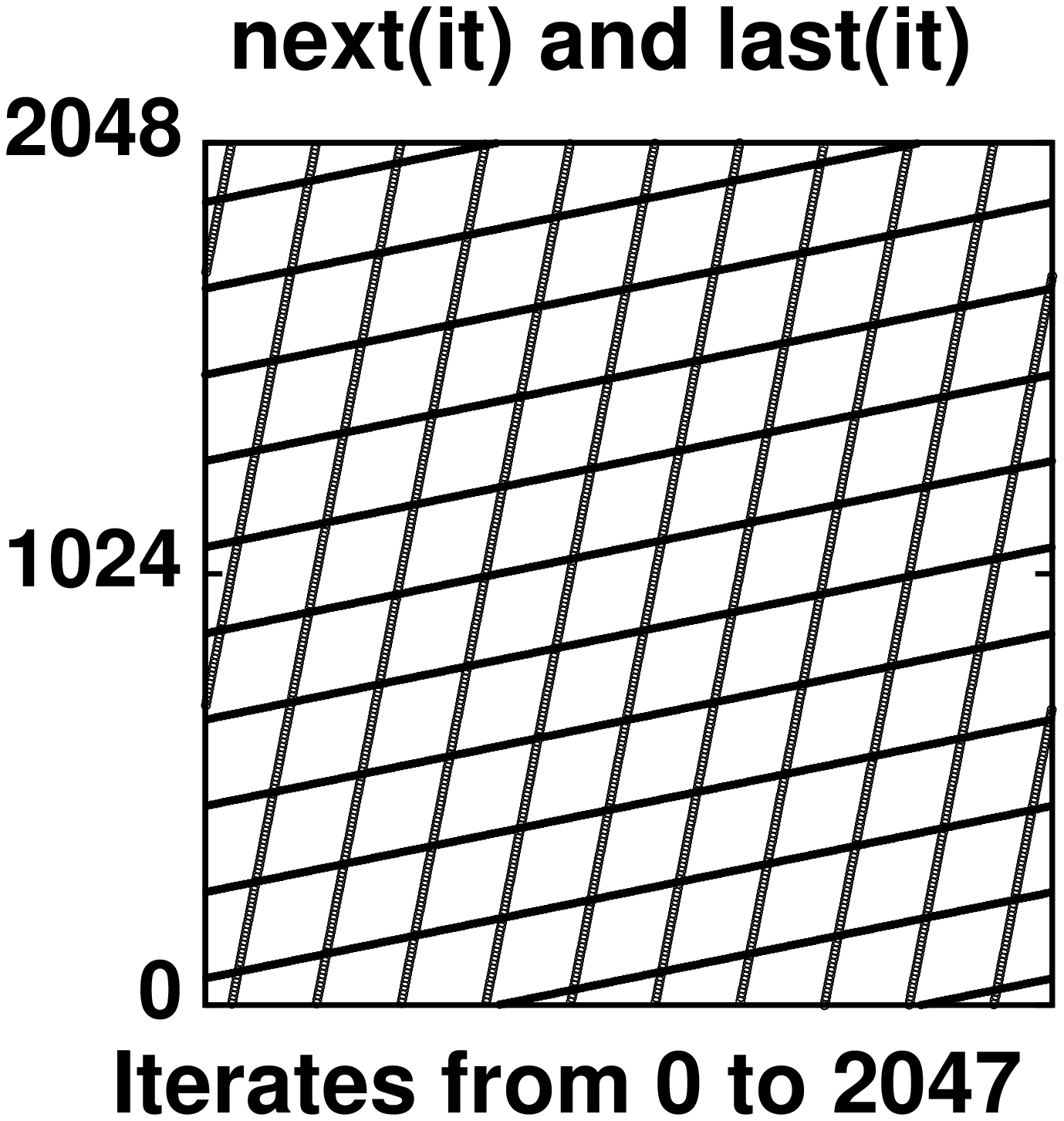}
\end{figure}

\vspace{5 mm}

\noindent
The {\tt last} function, likewise plotted in the Figure, is also single-valued.  It {\it appears} to
generate ten separate lines.  The values of {\tt last} for {\tt 1} $\le$ {\tt j}  $\le$ {\tt 11} are
just enough to indicate how this function resembles ten continuous lines :
$$
{\tt last} \rightarrow (1,1702),(2,1907),(3,64),(4,269),(5,474),
$$
$$
(6,679),(7,884),(8,1089),(9,1294),(10,1499),(11,1704) \ .
$$
Comparing the first and last entries shows that the apparent slope of the less-steep ``lines'' is
$(1704-1702)/(11-1) = (1/5)$ .  The reversed arrays of points have an apparent slope of (1/5) because
increasing {\tt j} by 10 is {\it usually} required in order to increase {\tt last} by 2 . These two
functions {\tt next} and {\tt last} go forward and backward, taking in the result of the most recent
iteration, in the range $[0,2047]$ , as the abscissa and returning, as the ordinate,  either the next or
the most recent integer.

\section{Our Challenge [ recently met! ] }

We viewed a simple, explicit time-reversible pair of generators analogous to {\tt next} and {\tt last}
above as desirable for stochastic computer simulations.  We have happily rewarded the first successful
FORTRAN algorithm generating the two-argument reversal of {\tt rund(intx,inty)} above with a cash
prize of 500 United States dollars, awarded to Federico Ricci-Tersenghi.  As was required, his solution
was demonstrated to work for the initial seeds {\tt intx = inty = 0} .  Here is a program based on his
solution:

\noindent
{\tt
      program federico \\
      parameter (items = 4194304) \\
      implicit integer(a-z) \\   
      dimension forwx(items),forwy(items),backx(items),backy(items) \\
\noindent
      intx = 0 \\
      inty = 0 \\
      do n = 1,items \\
      i = 1029*intx + 1731 \\
      j = i + 1029*inty + 507*intx - 1731 \\
      intx = mod(i,2048) \\
      j = j + (i - intx)/2048 \\
      inty = mod(j,2048) \\
      forwx(n) = intx \\
      forwy(n) = inty \\
      end do  \\

\noindent
      intx = 0 \\
      inty = 0 \\
      do n = 1,items \\
      oldx = mod(205*intx + 1497,2048) \\
      inty = inty + items - 1536*oldx - (1029*oldx + 1731 - intx)/2048 \\
      inty = mod(205*inty,2048) \\
      intx = oldx \\
      backx(n) = intx \\
      backy(n) = inty \\
      enddo \\
\noindent
      stop \\
      end \\
}

The seed variables forward and backward can be verified to satisfy the identities :
$$
\noindent
{\tt
forwx(i) = backx(items - i) \ ; \ forwy(i) = backy(items - i) \ .
}
$$

\section{Acknowledgments}

We are particularly grateful to the late Ian Snook and our colleague Niels Gr$\phi$nbech-Jensen for
calling the importance of stochastic algorithms to our attention, to Nathan Hoover for pointing out
the disallowed storage shortcut, and to Carl Dettmann for pointing out to us that the damped
harmonic oscillator {\it is} time-reversible in the sense put forward in Reference 2 .  We specially
thank Professor Ricci-Tersenghi for his prompt solution of our challenge.

\newpage

\section{Another Challenge?}

In 2014 we intend again to offer an Ian Snook Memorial Challenge Prize for solving an interesting
problem relevant to computational statistical mechanics.  Suggestions welcome!

\end{document}